\documentclass[sigconf,screen]{acmart}

\setcopyright{acmlicensed}
\acmDOI{10.1145/3663532.3664467}
\acmYear{2024}
\copyrightyear{2024}
\acmSubmissionID{fsews24fasegamesmain-id3-p}
\acmISBN{979-8-4007-0674-5/24/07}
\acmConference[FaSE4Games '24]{Proceedings of the 1st ACM International Workshop on Foundations of Applied Software Engineering for Games}{July 16, 2024}{Porto de Galinhas, Brazil}
\acmBooktitle{Proceedings of the 1st ACM International Workshop on Foundations of Applied Software Engineering for Games (FaSE4Games '24), July 16, 2024, Porto de Galinhas, Brazil}
\received{2024-03-28}
\received[accepted]{2024-04-26}

\usepackage{xspace}
\usepackage[capitalise]{cleveref}
\usepackage{hyperref}
\usepackage{subcaption}
\usepackage[
    colorinlistoftodos,prependcaption,textsize=tiny
]{todonotes}

\newcommand\definetool[2]{\newcommand{#1}{{\textsc{#2}}\xspace}}
\definetool{\Scratch}{Scratch}
\definetool{\Whisker}{Whisker}
\definetool{\Neat}{Neat}
\definetool{\Neatest}{Neatest}

\newcommand{\MeanCovCreateYourWorldNeatest}{79.52}
\newcommand{\MeanCovPokeClickerNeatest}{45.79}

\newcommand{\MeanCovCreateYourWorldNovelty}{82.73}
\newcommand{\MeanCovPokeClickerNovelty}{47.99}

\begin{document}

\title[Combining Neuroevolution with Novelty-Search for Testing Games]{Combining Neuroevolution with the Search for Novelty to Improve the Generation of Test Inputs for Games}

\author{Patric Feldmeier}
\email{patric.feldmeier@uni-passau.de}
\orcid{0000-0002-9509-7671}
\affiliation{%
  \institution{University of Passau}
  \country{Germany}
}

\author{Gordon Fraser}
\email{gordon.fraser@uni-passau.de}
\orcid{0000-0002-4364-6595}
\affiliation{%
  \institution{University of Passau}
  \country{Germany}
}


\begin{abstract}

  As games challenge traditional automated white-box test generators,
  the \Neatest approach generates test suites consisting of neural
  networks that exercise the source code by playing the games.
  \Neatest generates these neural networks using an evolutionary
  algorithm that is guided by an objective function targeting
  individual source code statements. This approach works well if the
  objective function provides sufficient guidance, but deceiving or
  complex fitness landscapes may inhibit the search.
  In this paper, we investigate whether the issue of challenging fitness
  landscapes can be addressed by promoting novel behaviours during the search.
  Our case study on two \Scratch games demonstrates that rewarding
  novel behaviours is a promising approach for overcoming challenging
  fitness landscapes, thus enabling future research on how to adapt
  the search algorithms to best use this information.

\end{abstract}




\maketitle

\section{Introduction}
\begin{figure}[!tbp]
    \centering
    \includegraphics[width=\columnwidth]{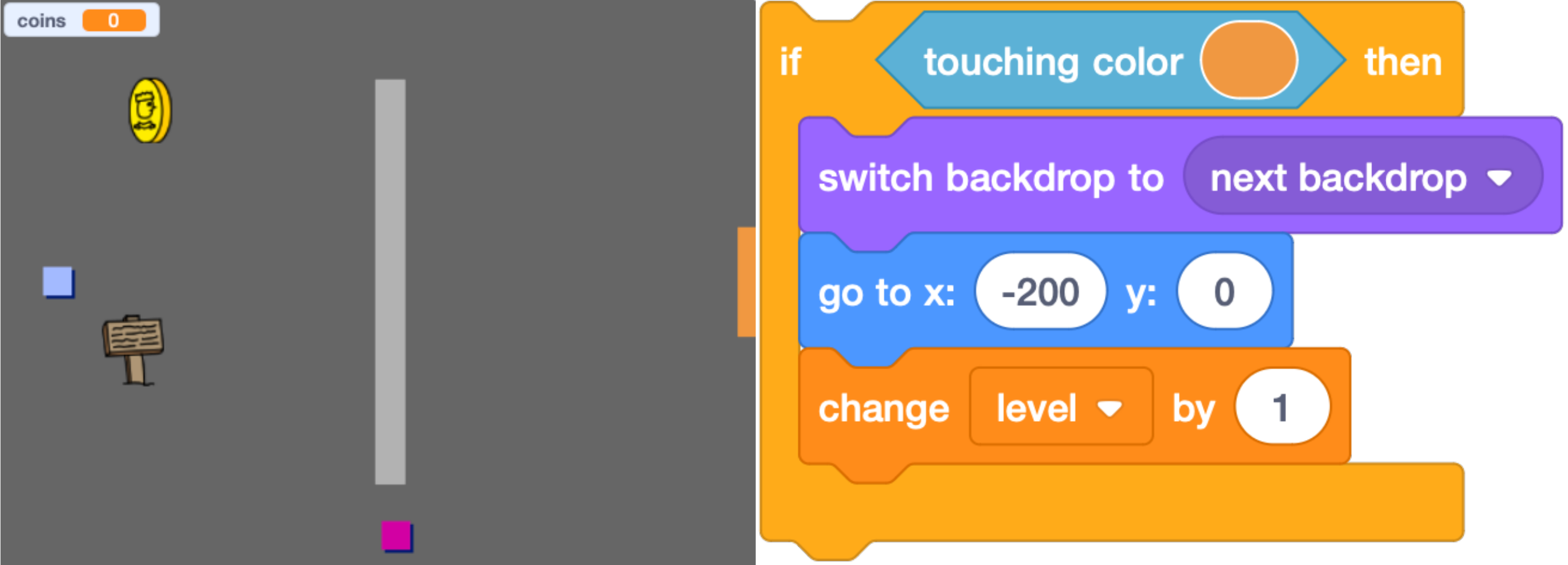}
    \caption{The \emph{CreateYourWorld} game together with the condition checking
    if the player has reached the next level by touching the orange portal
    on the right side of the level.}
    \label{fig:CreateYourWorld}
\end{figure}

In order to generate tests for games automatically, test generators
must be able to produce test input sequences capable of reaching a
wide variety of game states, such that the observed program behaviour
within these states can be checked. However, producing input sequences
for advanced program states is hard as games are designed to challenge
players with increasingly difficult tasks. Thus, test generators have
to learn to play them to create input sequences capable of reaching
advanced program states.

\Neatest tackles these challenges by generating test suites in which
every test case is represented by a neural
network~\cite{feldmeier2022neuroevolution}. Each of these networks
serves as a test input generator optimised by an evolutionary search
algorithm (i.e., neuroevolution) to reach a specific source code
statement reliably, regardless of challenging or randomised program
behaviour, by generating input actions for the given game dynamically
based on the current program state.
The search is guided by an objective function that measures the
distance of candidate test executions to program statements the search
has yet to reach~\cite{deiner2023automated}.

Although it is based on the source code, the objective function
includes aspects of program states. For instance, consider the
\emph{CreateYourWorld} \Scratch game shown
in~\cref{fig:CreateYourWorld}, where the player, represented by a blue
square, is tasked to manoeuvre through various levels by reaching
orange portals, such as located on the right side
in~\cref{fig:CreateYourWorld}. The corresponding \Scratch code shows
that the program determines whether the player is eligible to get to
the next level by evaluating if the player touches the portal's orange
colour. Whenever the player touches the orange portal, they are placed
into the next level by changing the scene's background image,
resetting the player's location to the left and increasing a variable
that keeps track of the current level. Since the if-condition serves
as a guard for reaching these three statements, an objective function
responsible for optimising test cases to reach one of those statements
will compute the \emph{Euclidean distance} between the player and the
target location~\cite{deiner2023automated}. However, achieving this
requires the player to temporarily increase the distance between the
player and the portal in order to pass the wall depicted in
grey. Although manoeuvring around the wall is necessary to reach the
portal, the search will penalise such behaviour as it leads to worse
fitness values.

\begin{figure}[!tbp]
    \centering
    \includegraphics[width=\columnwidth]{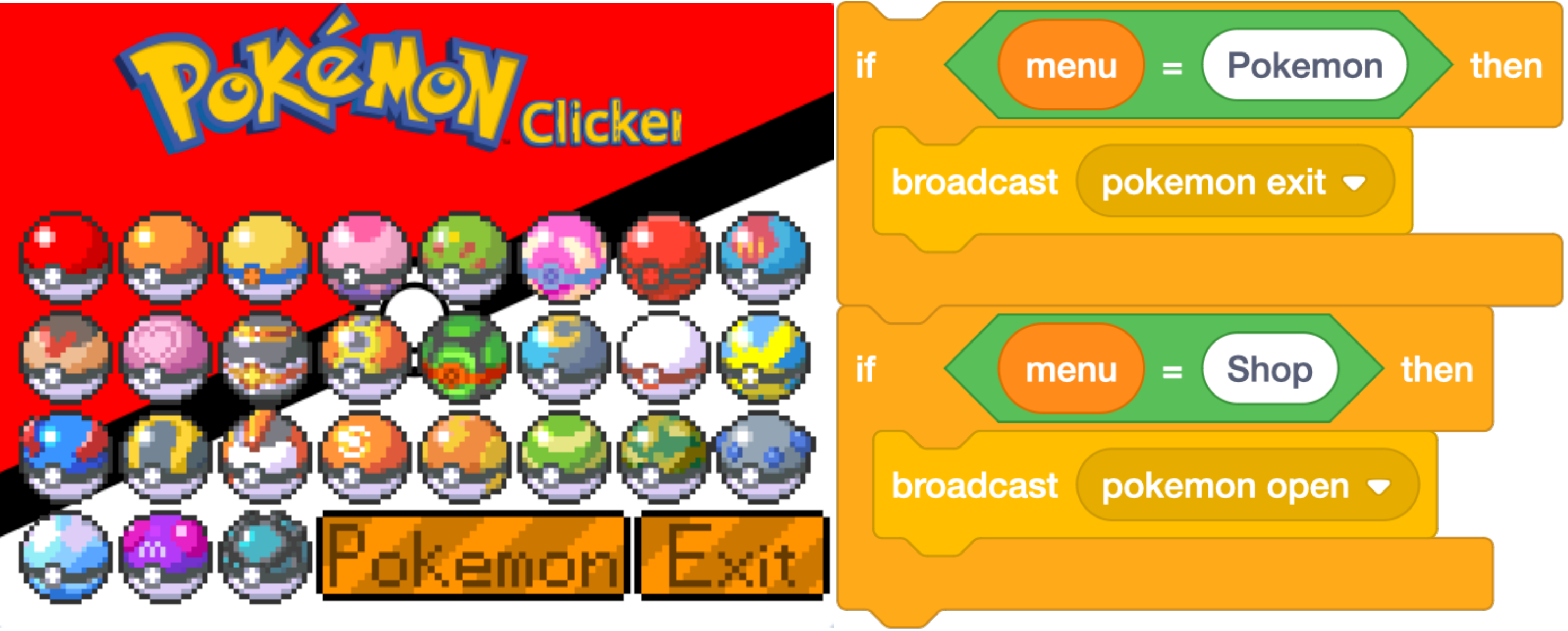}
    \caption{The \emph{PokéClicker} game together with two conditions
    that are evaluated when pressing the \emph{Pokémon} button and responsible
    for navigating the player to different menus.}
    \label{fig:PokeClicker}
\end{figure}

An orthogonal problem faced by objective functions is a lack of
guidance.  For instance, it is hard to define sensitive objective
functions that measure how close the test generator is to reaching
statements guarded by conditions involving non-numerical values such
as strings or events like button presses. Take as an example the
\emph{PokéClicker} game shown in~\cref{fig:PokeClicker}, in which
players can buy various upgrades by spending points they can earn
through repeatedly clicking on a Pokéball. Each of these menus
consists of two buttons that lead the player to the next or previous
menu screen, hosting a plethora of upgrades. As can be seen in the
code shown in \cref{fig:PokeClicker} that handles such button presses,
pressing the same button may lead to different game states indicated
by the name of the currently shown menu represented as a string. Since
the string distance in this case cannot provide meaningful guidance,
the search does not receive feedback on how close an execution came to
reaching statements that are guarded by respective conditions.
Fitness landscapes that lack guidance or are deceptive harm the search
progress and may even render the search unable to find satisfying
solutions~\cite{lehman2008exploiting,christensen2007solving,langdon1998ants}.

In order to overcome deceptive fitness landscapes and those lacking
guidance, previous work has suggested abandoning or combining
objective functions that measure the performance within a problem
domain and instead reward agents that interact with the environment in
novel, previously unseen
ways~\cite{lehman2010efficiently,risi2009how}. Thus, an agent is not
penalised for temporarily increasing the distance to a target location
but is instead rewarded for exploring novel areas in the scene and
experimenting with UI elements in different ways.

In this paper, we conduct an initial exploration of the integration of
novelty search in neuroevolution-based test generation for games by
encouraging the optimisation of novel behaviours in \Neatest. To this
end, we derive the behaviour of an agent based on the final state
reached after an agent interacted with the game and integrate novelty
as a secondary fitness criterion in \Neatest. In detail, the
contributions of this paper are as follows:
\begin{itemize}
	\item We propose integrating novelty search into the test generation
	for games to cope with challenging fitness landscapes.
	\item We implement the proposed approach as an extension to the publicly available game testing framework \Neatest.
	\item In a case study on two \Scratch games, we empirically
          demonstrate how novelty can help to overcome challenging
          fitness landscapes.
\end{itemize}


\section{Search-based Software Testing For Games via Neuroevolution}
\label{section:sbst}

In this paper we investigate combining \Neatest, a test
generator for games that uses neuroevolution as a search-based
software testing method, with an approach that promotes novel
behaviours of evolved solutions.  Search-based software testing
(SBST)~\cite{mcminn2004search} uses meta-heuristic search algorithms,
such as evolutionary algorithms, to generate tests for a given
application. Based on the chosen representation of viable solutions,
search algorithms may be applied to a variety of software testing
challenges, such as generating sequences of method
calls~\cite{fraser2011evosuite, paolo2004evolutionary,
  baresi2010testful}, synthesising inputs for testing GUI
applications~\cite{sell2019empirical, gross2012search,
  mao2016sapienz}, and optimising calls to REST
services~\cite{arcuri2019restful}.

To succeed in this endeavour, the search algorithms must be guided by
an objective function $f$ that evaluates how close a generated test is
to reaching the optimisation goal. For example, if test cases are to
be optimised to reach a specific program statement~$s$, the objective
function must determine how close a generated test input sequence
comes to reaching the target statement. The target statement counts as
covered by the test case $t$ if $f_s(t) = 0$ after executing the test
case on the program. Otherwise, if the test case has not covered the
target ($f_s(t) > 0$), it is further optimised via search algorithms
by minimising the objective function until the statement is covered. A
commonly used fitness function is defined by a linear combination of
the approach level $A$ and the branch distance $B$. The approach level
measures how close an executed test case in the control dependence
graph (CDG)~\cite{CDG} is from executing the control location on which
the target branch depends. The branch distance is evaluated at the
control location where a path through the program is taken that makes
the execution of $s$ impossible, and heuristically computes how far
the predicate of the control location is from obtaining the opposite
value that paves the way towards reaching $s$. For example, suppose
that a predicate evaluated to \emph{true} during the execution of a
test case $t$, the branch distance computes the distance towards
evaluating the same predicate to \emph{false}.

Based on these two metrics, we define the objective function
\label{eq:statementFitness} $f_s(t) = A_p(t) + \omega(B_p(t))$
that measures the distance of an executed test case $t$ to the targeted statement $s$
by applying the normalisation function $\omega$ to the branch distance $B$. This
normalisation function $w:\mathbb{R}^+\rightarrow [0,1]$ maps the branch
distance to values lower than 1 in order to guarantee that the approach level
defined as an integer is weighted more important than the branch distance~\cite{arcuri2010normalise}.

We consider games written in the \Scratch~\cite{maloney2010scratch}
programming language, which is a block-based programming language that
helps young students in learning to program. Each \Scratch program
consists of \emph{Sprites} representing figures in the game, similar
to \emph{Game Objects} in \emph{Unity}~\cite{haas2014history}, and a
\emph{Stage} that sets the scene on top of which the \emph{Sprites}
act. The program logic is implemented by arranging and combining
blocks that represent programming statements into scripts. These
scripts are added to \emph{Sprites} to model their behaviour or to the
\emph{Stage} to implement general properties of the scene.
\Scratch programs can be tested using the
\Whisker~\cite{stahlbauer2019testing} framework. \Whisker maps Boolean
predicates such as the one shown in the code snippet of
\cref{fig:CreateYourWorld} to branch distance measurements based on
the underlying program state (e.g., the distance between the current
sprite and the closest pixel with the chosen target
colour)~\cite{deiner2023automated}.

\Neatest~\cite{feldmeier2022neuroevolution} extends \Whisker to tackle
the many challenges of game testing by generating dynamic test suites.
Within these test suites, each test case corresponds to a direct policy neural
network agent optimised to reach a targeted
program state by generating test-inputs for the program dynamically
based on the current state of the program. To guarantee that the
optimised network agents are robust against program
randomisation, \Neatest only counts a target statement $s$ as
covered if the agent passes a robustness check that validates weather the agent is able to
reach the targeted statement repeatedly in several randomised
program executions. If a statement is covered reliably and passes the
robustness check, the search proceeds with optimising
networks for the next statement $s$ using the objective function
$f_s(t)$. This objective function is independent of the game being
tested, which allows \Neatest to implicitly learn to win or lose games
in different ways without requiring any domain-specific knowledge, such as the high
score achieved or the time survived. Target statements are
selected by querying the (CDG)~\cite{CDG} for statements that are
direct children of already covered program statements, allowing
\Neatest to apply a variation of \emph{Curriculum
  Learning}~\cite{bengio2009curriculum} where the game is explored
iteratively by seeding initial generations with prior solutions.

For a chosen target statement, \Neatest optimises networks via the
neuroevolution algorithm \Neat~\cite{stanley2002evolving} that applies
evolutionary search operators over many generations in order to explore the
search space of viable solutions. Evolutionary algorithms are inspired by
Darwinian evolution
as new generations are evolved by first selecting parents
based on their achieved fitness value that is derived from a pre-defined fitness
function such as the one defined above. The
selected parents are then evolved using mutation and crossover operators. Within the
\Neat algorithm, mutations introduce probabilistic variation by extending a network’s
topological structure or by changing attributes of existing genes such as the
weights of a network. Crossover forms a single child from two parents by
combining the genes of both parents. Selecting parents only based on
their performance within their problem domain, might stall the search progress
since in some scenarios temporarily decreasing the achieved fitness value might
be a necessary step towards reaching a satisfying solution~\cite{lehman2008exploiting,christensen2007solving,langdon1998ants}.

\section{Combining Test Generation for Games with the Search for Novelty}
\label{sec:NovelTest}

Novelty search algorithms behave like other evolutionary search algorithms~\cite{arcuri2018test,fraser2011evosuite} as they
repeatedly apply selection, mutation and crossover operations on their
individuals to form new generations. However, instead of using an objective
 function that measures how close an individual is
to finding a solution, novelty search algorithms evaluate individuals based on
the novelty of their behaviour~\cite{lehman2010efficiently,risi2009how}.
To this end, a novelty metric is defined that
operates within the space of feasible behaviours to measure the distance between
an individual's behaviour and already observed behaviours collected in an
archive. The novelty score $n(t)$ of an individual $t$ is commonly computed using the
$k$-nearest neighbours algorithm $n(t) = 1/k \sum^k_{i=0} d(x, \mu_i)$,
with $\mu_i$ representing the $i$-th nearest neighbour of $t$ with respect to the
distance metric $d:\mathbb{R}^+\rightarrow [0,\infty]$.
This metric aims to capture behavioural differences between two
individuals in the domain-specific behaviour space. For instance, in a two-dimensional maze navigation task, the
behaviour space might be defined by all coordinates that are reachable
within the maze. Since high novelty scores point to behaviours in
 that have not been explored
thoroughly, the resulting novelty-based fitness function expresses a gradient
toward behavioural difference that puts constant pressure on discovering
novel solutions and avoids deceptive fitness landscapes~\cite{lehman2008exploiting,christensen2007solving,langdon1998ants}.

There are many different possibilities of how to integrate novelty metrics into search algorithms. For our initial investigation, we integrate novelty in \Neatest as a secondary fitness criterion such that the novelty 
 acts as a tiebreaker between individuals that have performed
equally well according to the objective function. For example, during selection individuals with identical fitness values are ranked based on their novelty scores.
The behaviour of an
evolved neural network is derived from the program state reached after executing
the network agent within the problem domain.

Since \Neatest aims to generalise to any game regardless of its genre, we avoid extracting
genre-specific game states such as the coordinates in a maze navigation task.
Instead, we extract the same features from the game \Neatest already uses
as an input signal to the networks. Although we define the state features from
which observed behaviour is defined from \Scratch games in this work, this
approach also generalises to other programming environments. State features are
extracted by iterating over all \emph{visible} figures on the game screen while
deriving the following attributes from corresponding \emph{Sprites} that are
bounded by the \Scratch environment~\cite{maloney2010scratch} and normalised
into the range $[0,1]$:
\begin{itemize}
\item \textbf{Position} defined by 2-dim coordinates $a \in [-240, 240]$ and $b\in[-180, 180]$ on a 2-dim game canvas.
\item \textbf{Heading direction} defined by an angle $\alpha \in [-180, 180]$.
\item \textbf{Costume} of a figure defined by the index $i \in N$ over the list of available costumes if the figure changes its appearance.
\item \textbf{Size} of a figure defined by the size (\%) of the selected costume. 
\item \textbf{Private variable} values $v \in \mathbb{R}$ for each numeric variable.
\item \textbf{Distance} $d\in[-600, 600]$ to a sprite or colour if the figure contains listeners for touching other sprites or colours.
\end{itemize}
Besides attributes specific to visible figures, we also collect global variables
that host numeric values ($v \in \mathbb{R}$) and the mouse position
($a \in [-240, 240]$, $b\in[-180, 180]$) if the game contains code that
listens to the position of the mouse.

In order to calculate the distance $d$ between two observed behaviours, we
compute the \emph{cosine similarity} over both behaviours after arranging them
into feature vectors in which every index corresponds to a specific
state feature~\cite{hao2014puma}. The output space of the \emph{cosine
similarity} function $cos$ is restricted to the interval $[-1,1]$, where two
parallel vectors have a similarity of 1, two orthogonal
vectors a similarity of 0, and two opposite vectors a similarity of~-1. However,
for our application scenario, we
normalise the obtained similarity score $d_{cos}$ via $d_{cos}=(d_{cos} + 1) / 2$. This normalisation step
outputs a similarity score of 0 if the vectors represent opposite states, 0.5
if they are orthogonal to each other and 1 if they are the same. Finally, since
\Neatest is modelled as a maximisation task, and we want to reward low
degrees of similarity, we invert the obtained similarity score by subtracting it
from one, which results in the formula $d_{cos} = 1 - ((cos(s1, s2) + 1) / 2)$ for computing the similarity
distance between two observed game states $s_1$ and $s_2$.


\section{Case Study}

Our case study analyses whether the
neuroevolution-based generation of test cases benefits from promoting novel
behaviours. To this end, we extend \Neatest, which is part of the open-source
\Whisker testing framework~\cite{deiner2023automated} with a novelty score
integrated into the test generator as a secondary fitness criterion as
explained in~\cref{sec:NovelTest}. The effect of adding novelty-rewarding mechanisms to the search is evaluated by comparing the achieved
coverage of the default \Neatest algorithm (\emph{Fitness}) against our proposed
approach (\emph{Fitness+Novelty}). In addition, we compute the \emph{Vargha
and Delaney} effect size ($A_{12}$)~\cite{vargha2000critique} between the two
approaches and determine statistical significance based on
the \emph{Mann-Whitney-U} test~\cite{Mann-Whitney-U} using a significance
threshold of 0.05. To account for randomisation inherent to
neuroevolution, both algorithm configurations are executed ten times against our
 two case study \Scratch games.

\begin{figure}[!tbp]
    \centering
    \includegraphics[width=\columnwidth]{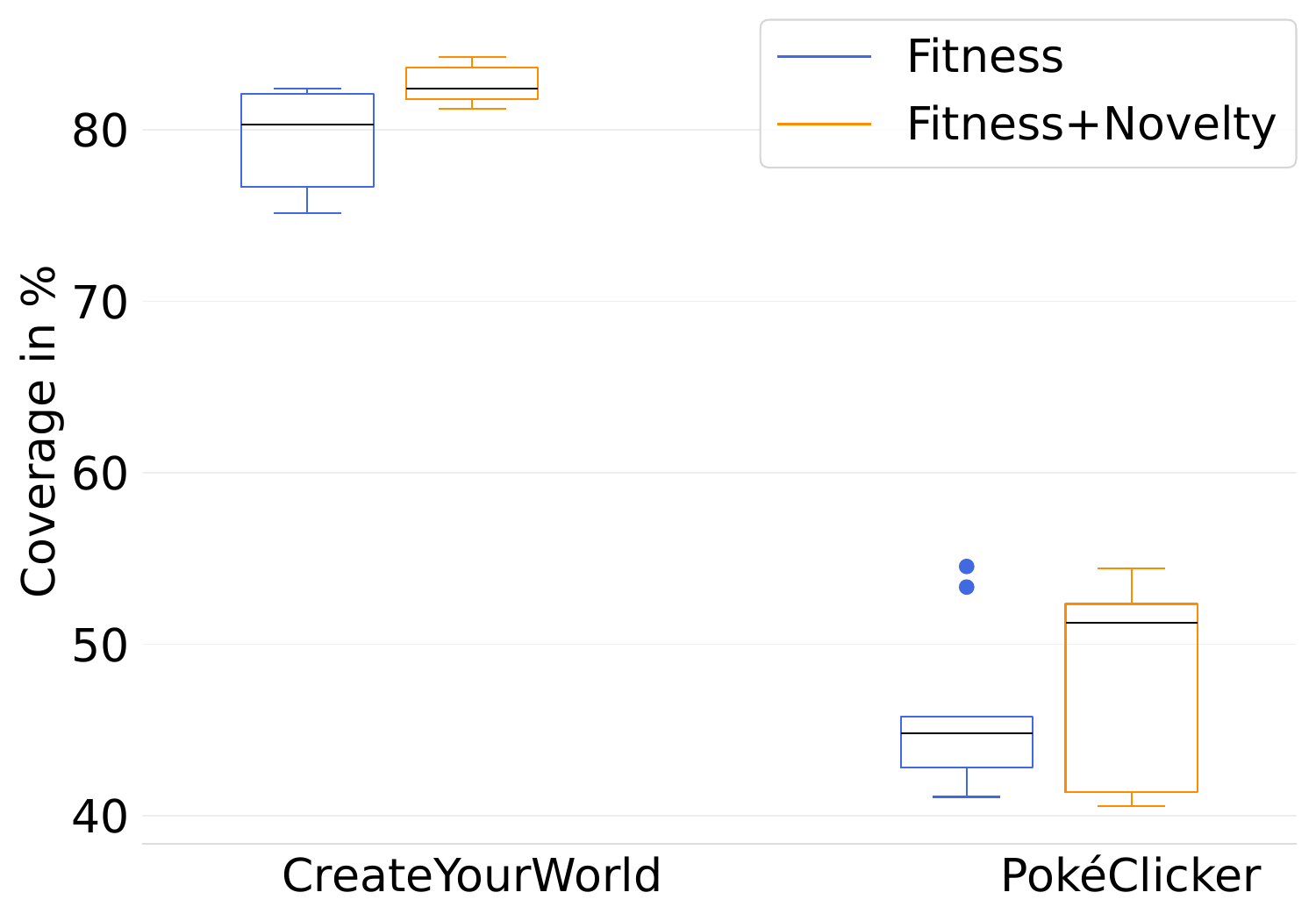}
    \caption{Achieved coverage of \Neatest and the proposed novelty approach over both dataset games.}
    \label{fig:Box}
\end{figure}

Both games correspond to a different genre of which we hypothesise that
\Neatest might suffer from  challenging fitness landscapes: maze
navigation (\emph{CreateYourWorld}) and clicker games (\emph{PokéClicker}). \emph{CreateYourWorld} was extracted from an
introductory tutorial\footnote{[March2024]: \url{https://codeclub.org/en/}}
while \emph{PokéClicker} was gathered by searching for the clicker genre
on the \Scratch website\footnote{[March 2024]:
\url{https://scratch.mit.edu}} and sorting the results by popularity. Although
the case study games are implemented in \Scratch, they
nevertheless represent games that may also be found online as web games.

All experiments were conducted on a computing cluster consisting of AMD Epic
7443P CPU cores with a clock frequency of 2.85GHz. In order to reduce
the time required for evaluating network agents in the problem domain, we make
use of \Whisker's test execution acceleration and update the game state as fast
as the employed CPU cores can process action events that are sent to the game by
the test generator. Similar to previous research, we only count a statement as
covered if the test generator is able to reach the same target in ten
randomised program
executions~\cite{feldmeier2022neuroevolution,feldmeier2023learning}. Every
network is allowed to play the game until it reaches a game over state, wins the
game or has played the game for five seconds, which corresponds to a
much higher game time due to the employed program acceleration. In our
experiments, \Neatest evolves
a population of 150 networks using speciation coefficients, as well as mutation and crossover probabilities that have
shown to work well in the \Scratch domain, according to previous
work~\cite{feldmeier2023learning}. We use a search
duration of ten hours since challenging landscapes may only be encountered after
some time has passed. To avoid penalising novel
discoveries, we add observed behaviours
with a probability of 0.1 to the behaviour archive regardless of their novelty
score. This probability is chosen together with a $k$ value of 15 for the $k$
nearest neighbour algorithm as it has shown to produce promising results in previous
work~\cite{lehman2008exploiting,jackson2019novelty}.

\subsection{Threats to Validity}
The evaluation uses two \Scratch games for which we hypothesised that
the search would benefit from promoting novelty in order to evaluate
the effect of integrating novelty strategies into \Neatest. In future
work, we seek to extend the dataset by sampling games without this
bias to confirm that the results generalise. We mitigate the effects
of randomisation inherent to the neuroevolution algorithm by repeating
every experiment ten times and determining statistical significance
using the \emph{Mann-Whitney-U} test~\cite{Mann-Whitney-U}.  We use
block coverage as a proxy to measure the effectiveness of an evolved
test suite. However, block coverage must be treated with caution in
\Scratch as high coverage values may already be achieved by simply
starting the game.

\subsection{Results}
As shown in \cref{fig:Box}, our study on two games involving challenging
fitness landscapes reveals that \Neatest benefits from integrating novelty search
in the neuroevolution algorithm with an increase in
average program coverage to \MeanCovCreateYourWorldNovelty\% and
\MeanCovPokeClickerNovelty\% from \MeanCovCreateYourWorldNeatest\% and
\MeanCovPokeClickerNeatest\% for the games \emph{CreateYourWorld}
and \emph{PokéClicker}, respectively.

\begin{figure}[!tbp]
    \centering
    \includegraphics[width=\columnwidth]{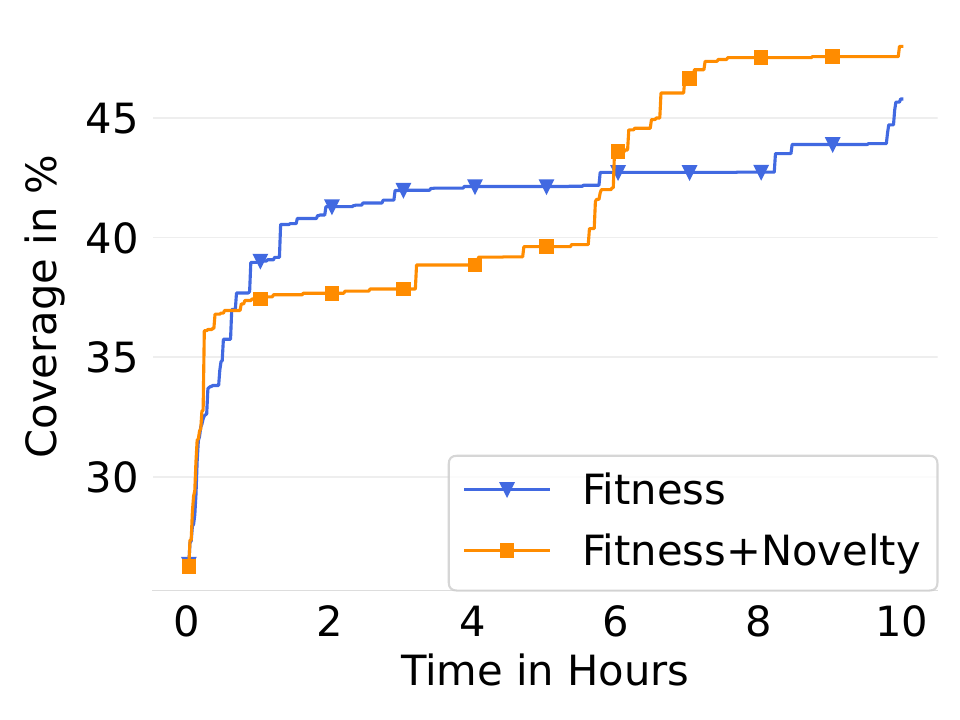}
    \caption{Achieved coverage over time of \Neatest and the proposed novelty approach for the PokéClicker game.}
    \label{fig:Time-PokeClicker}
\end{figure}


In the \emph{PokéClicker} game shown in~\cref{fig:PokeClicker}, high program coverage
may be achieved by navigating through the many menus of this game effectively by
pressing buttons in these menus at different game states. However, since many
conditions in this game compare two strings against each other, the objective
function cannot give the search any guidance towards reaching statements
guarded by the respective conditions. Thus, most individuals get assigned the
same fitness value regardless of how they interact with the game. As indicated
by the coverage plateau in~\cref{fig:Time-PokeClicker}, which \Neatest encounters after
around one hour, this lack of guidance affects the search negatively. In
contrast, analysing the achieved coverage over time for
the novelty approach, we can observe that adding novelty as a secondary fitness criterion
helps overcome this plateau by selecting promising parents to evolve in the
presence of many similarly rated individuals. Especially after around five to six hours, we can observe an enormous
increase in the achieved program coverage. This is very likely a point within the search at which the
selection pressure increases dramatically toward novel behaviours due to the
behaviour archive being filled with many common behaviours.

\begin{figure}[!tbp]
    \centering
    \includegraphics[width=\columnwidth]{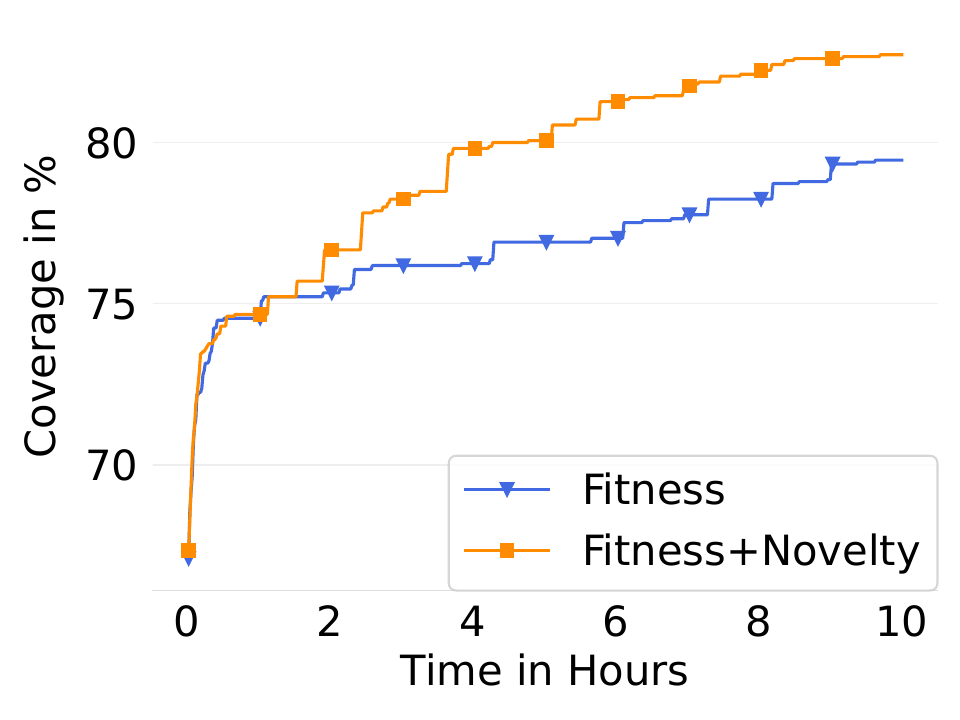}
    \caption{Achieved coverage over time of \Neatest and the proposed novelty approach for the CreateYourWorld game.}
    \label{fig:Time-CreateYourWorld}
\end{figure}

A significant ($p$ < 0.01) increase in coverage from \MeanCovCreateYourWorldNeatest\% to
\MeanCovCreateYourWorldNovelty\% and an effect size of 0.85 can be observed
for the \emph{CreateYourWorld} game. To shed more light on this observation, we analysed
how often each algorithm configuration was able to reach the next level in this
maze navigation game by touching the orange portal depicted in \cref{fig:CreateYourWorld}. While
\Neatest overcomes the grey wall separating the player and the orange door
only in 7/10 cases, our proposed
novelty-based approach manages to reach the next level in every experiment
repetition. As depicted in \cref{fig:Time-CreateYourWorld}, the novelty search approach reaches
the next level relatively early in the search, after around two hours,
allowing the test generator to explore more parts of the game and reach much
higher program coverage.

Overall, our case study demonstrates that integrating novelty search
into the optimisation of tests for games is a promising approach to
overcome deceptive fitness landscapes encountered in maze navigation
tasks. Furthermore, adding the novelty score as an additional fitness
criterion has been shown to assist the search in scenarios with poor
optimisation gradients, as frequently encountered in games that
involve many menus, such as clicker games.

\section{Related Work}

To this day, most video games are tested
manually~\cite{desurvire2004using,ferre2009playability} or use
semi-automated testing
approaches~\cite{schaefer2013crushinator,cho2010online,smith2009computational}. However,
since games are commonly built incrementally, both approaches are not
only time-consuming but also expensive and error-prone, as testers
have to re-validate the same program in updated program versions
repeatedly~\cite{schultz2016game,politowski2021survey}.  Various
approaches for generating tests automatically have been proposed, such
as combining reinforcement learning with computer vision
techniques~\cite{paduraru2021automated}, employing evolutionary search
to find simulation traces~\cite{casamayor2022bug} or combining
reinforcement learning with evolutionary algorithms and
multi-objective optimisation~\cite{zheng2019wuji}. To tackle the
challenges of heavy program randomisation inherent to games, \Neatest
evolves test cases in the form of neural networks that produce input
sequences dynamically based on the current state of the
program~\cite{feldmeier2022neuroevolution}. \Neatest has been
previously improved using
gradient descent~\cite{feldmeier2023learning} for mutation.  In this
work, we extend \Neatest with the search for novel solutions to cope
with challenging fitness landscapes.

Novelty search has previously been combined with neuroevolution to
escape deceptive fitness landscapes in various
tasks~\cite{lehman2010efficiently,
  lehman2008exploiting,risi2009how}. Since abandoning the actual
objective may be too harsh in many scenarios, previous research has
investigated combining novelty-based and objective-based fitness
functions via multi-objective evolutionary algorithms using
\emph{Pareto dominance}~\cite{voorneveld2003characterization} to sort
the population~\cite{mouret2011novelty}. While we aim to promote the
search for novel solutions to reach a wide variety of states within a
game for testing purposes, previous work has used novelty search to
reach a specific goal within an environment, like winning a game or
navigating through a maze. To the best of our knowledge, applying
novelty search for testing games has not been explored previously.

\section{Conclusions}
Test generators for games must be capable of producing test inputs
that explore many areas of a game while also solving game objectives
in order to reach advanced program states. \Neatest tackles the
challenges of generating tests for games by optimising test cases
using an objective function that minimises the distance towards
reaching yet uncovered program statements. Since this objective
function might not produce good guidance for the search in specific
scenarios, this work explores extending \Neatest with a secondary
fitness criterion that rewards novel test behaviour. A case study on
two \Scratch games demonstrates that novelty search may be a promising
approach to overcome challenging fitness landscapes, as it can help to
increase the achieved program coverage.

In the future, we aim to evaluate different approaches of extending \Neatest with
novelty search, such as casting the problem of evolving diverse suites for games
into a multi-objective optimisation task~\cite{mouret2011novelty}. Furthermore, we envision
using novelty search to improve the test generator's fault detection
capability by evolving multiple test cases for the same program
state that exhibit different behaviours and thus test the same state in different
ways.

\begin{acks}
  This work is supported by \grantsponsor{FR 2955/3-1}{DFG project FR2955/3-1
  “TENDER-BLOCK: Testing, Debugging, and Repairing Block-based
  Programs”}{https://gepris.dfg.de/gepris/projekt/418126274}. The authors are
  responsible for this publication's content. 
\end{acks}

\balance

\bibliographystyle{ACM-Reference-Format}
\bibliography{library}

\end{document}